\documentclass{ieeeaccess}
\usepackage{cite}
\usepackage{amsmath,amssymb,amsfonts}
\usepackage{algorithmic}
\usepackage{graphicx}
\usepackage{textcomp}
\usepackage{placeins}
\usepackage{hyperref}
\usepackage{xspace}
\usepackage{booktabs}
\usepackage{multirow}
\usepackage{makecell}
\usepackage{float}
\usepackage{algorithm}
\usepackage{subfig}
\usepackage{dsfont}
\usepackage{bbm}
\usepackage[nameinlink,capitalize]{cleveref}

\def\BibTeX{{\rm B\kern-.05em{\sc i\kern-.025em b}\kern-.08em 
    T\kern-.1667em\lower.7ex\hbox{E}\kern-.125emX}}

\newcommand{\citep}{\cite}
\usepackage{multibib}
\newcites{AP}{Additional References}

\graphicspath{{images//}}
\DeclareGraphicsExtensions{.eps,.png,.pdf,.jpg,.tif,.tiff,.ps}

\definecolor{navy}{rgb}{0.1, 0.1, 0.8}
\definecolor{gray}{rgb}{0.4, 0.4, 0.4}
\definecolor{olive}{rgb}{0.1, 0.5, 0.1}
\definecolor{ruby}{rgb}{0.8, 0.1, 0.3}
\definecolor{darkpastelgreen}{rgb}{0.01, 0.75, 0.24}
\definecolor{celestialblue}{rgb}{0.29, 0.59, 0.82}
\definecolor{coral}{rgb}{1.0, 0.5, 0.31}
\definecolor{Goldenrod}{rgb}{0.8,0.8,0}

\usepackage{soul}
\setstcolor{red}
\newcommand{\eat}[1]{}

\DeclareMathOperator{\Real}{\mathbb{R}}

\DeclareMathOperator{\His}{\mathcal{H}}

\newcommand{\Netsense}{{\textsc{{Net\-Sense}}}\xspace}
\newcommand{\Nethealth}{{\textsc{{Net\-Health}}}\xspace}

\newcommand{\titlename}{Predicting Relationship Labels and Individual Personality Traits from Telecommunication History in Social Networks using Hawkes Processes}

\begin{document}
\history{Date of publication xxxx 00, 0000, date of current version xxxx 00, 0000.}
\doi{10.1109/ACCESS.2017.DOI}

\title{\titlename}

\author{\uppercase{Mateusz Nurek}\authorrefmark{1}, \uppercase{Rados{\L}aw Michalski}\authorrefmark{1}, \uppercase{Omar Lizardo}\authorrefmark{2}, and \uppercase{Marian-Andrei Rizoiu}\authorrefmark{3}}

\address[1]{Wroc{\l}aw University of Science and Technology, Department of Artificial Intelligence, Wroc{\l}aw, Poland}
\address[2]{University of California Los Angeles, Department of Sociology, Los Angeles, U.S.}
\address[3]{University of Technology Sydney, The Data Science Institute, Sydney, Australia}

\tfootnote{This work has been partially funded by the National Science Centre, Poland, grant number 2021/41/B/HS6/02798, 
and the Commonwealth of Australia (represented by the Defence Science and Technology Group) through a Defence Science Partnerships Agreement.
The NetSense study was funded by the U.S. National Science Foundation (CISE) \#0968529.
The NetHealth study was funded by the U.S. National Institutes of Health (NIH) \#R01HL117757.}

\markboth
{Nurek \headeretal: Predicting Relationship Labels and Individual Personality Traits from Telecommunication History}
{Nurek \headeretal: Predicting Relationship Labels and Individual Personality Traits from Telecommunication History}

\corresp{Corresponding author: Mateusz Nurek (e-mail: mateusz.nurek@pwr.edu.pl).}

\begin{abstract}
Mobile phones contain a wealth of private information, so we try to keep them secure.
We provide large-scale evidence that the psychological profiles of individuals and their relations with their peers can be predicted from seemingly anonymous communication traces -- calling and texting logs that service providers routinely collect.
Based on two extensive longitudinal studies containing more than 900 college students, we use point process modeling to describe communication patterns.
We automatically predict the peer relationship type and temporal dynamics, and assess user personality based on the modeling.
For some personality traits, the results are comparable to the gold-standard performances obtained from survey self-report data.
Findings illustrate how information usually residing outside the control of individuals can be used to reconstruct sensitive information.

\end{abstract}


\titlepgskip=-15pt

\maketitle


\hypertarget{sec:introduction}{\section{Introduction}}

It is a well-known fact that we give away personal information whenever we act and interact online~\cite{Settanni2018}.
This has been repeatedly shown for online social media platforms such as Facebook~\cite{Kosinski2013,Youyou2015}, Twitter~\cite{Mao2011}, and even knowledge creation sites like Wikipedia~\cite{Rizoiu2016}.
More specifically, people's personality profiles can been predicted from the activity traces they leave behind in personal websites~\cite{Marcus2006}, blogs~\cite{Yarkoni2010}, Twitter messages~\cite{Kern2020} or Facebook profiles~\cite{Kosinski2013}.

Recently, Stachl et al~\cite{Stachl2020} showed that personality profiles could be estimated using information collected from users' smartphones. This includes digital traces of communication and social behavior, music consumption, app usage, mobility, overall phone activity, and day- and nighttime activity.
However, obtaining such data requires access to the user's social media activity or phones, which might provide specific degrees of safety to the user who controls the privacy settings.
In this work, we investigate what can be learned about users using data normally outside their control, e.g., from communication patterns with their peers. We find that even seemingly anonymous information can be used to predicts, with a high level of accuracy both individual characteristics and properties of their social relationships. 

This paper addresses three open questions concerning modeling and learning from call and text data traces.
The first question relates to modeling the communication patterns between individuals.
It is known that human communication is bursty and that it exhibits a long-tail distribution of inter-event times~\cite{Malmgren2008,Jo2012, urena2020estimating}.
Hawkes point processes have been successfully applied to model other bursty phenomena, such as information diffusion~\cite{Mishra2016FeaturePrediction} or neuronal firing patterns in the human brain~\cite{Johnson1996}.
The question is \textbf{can we model the call patterns between individuals using Hawkes point processes?}
The second question relates to learning about relationships between individuals using call patterns.
Hawkes processes fitted on event series have been shown helpful in predicting the final popularity of online items~\cite{Mishra2016FeaturePrediction}, and even differentiating between controversial and authoritative news sources~\cite{Kong2020a}.
The question is, therefore, \textbf{can we differentiate the relationship type between two individuals, given their call and texting timing series fitted using a Hawkes model?}
The third question relates to inferring users' personality traits from their call series.
While there exist several prior work learning user personality traits using mobile phone data~\cite{Harari2019,Stachl2020,Monsted2018}, these usually rely on behavioral information collected via sensor and log data from smartphones, which requires access to the users' phones.
The question is \textbf{can we use the Hawkes model outputs, fitted on the call series involving a given user, to infer their personality traits?}

{\bf The \Netsense and \Nethealth Studies}. We address the above open questions using two large datasets -- \Netsense and \Nethealth~ -- obtained from the NetSense and {Net\-Health} longitudinal social network studies conducted on a student population at the University of Notre Dame~\cite{striegel2013lessons, purta2016experiences}.
The datasets contain digital trace timestamped records of the calling and texting information for about 200 and 700 students, respectively, who also filled in periodic surveys administered at the start and end of each semester for two and three years in each respective study.
We answer the first question by fitting the parameters of a Hawkes point process to the series of communication events -- phone calls or texts -- occurring between each pair of students.
Using goodness of fit tests, we show that Hawkes with a power-law decaying kernel function generalizes better to unseen data than the exponential kernel.

To answer the second question, we use the student surveys in which they
provided labels for their most salient (top-20) contacts in periodic ego-network surveys (e.g., \emph{family}, \emph{friend}, \emph{significant other} and the like).
Based on the labels given to relationships by students in subsequent surveys, we first split the pairwise relations in both datasets into six categories: family-relaxing, family-stable, friendship-relaxing, friendship-stable, {friend\-ship}-strengthening and romantic-relaxing.
Next, we characterize each pairwise relationship in our dataset using the Hawkes model's fitted parameters and secondary quantities.
We show that using the Hawkes descriptors, off-the-shelf classifiers significantly outperform autoregressive (ARIMA \cite{Makridakis1997}) baselines, achieving a macro F1-score of $0.24$ for \Netsense and $0.22$ for \Nethealth), with the best-identified categories being family-stable, friendship-stable, and friendship-relaxing.
We obtain the best prediction results when concatenating the autoregressive and Hawkes features (F1-score of $0.27$ and $0.23$ on \Netsense and \Nethealth, respectively).

We answer the third question by applying a novel joint modeling of Hawkes processes~\cite{Kong2020a} to model jointly the call series of each student participating in the \Netsense and \Nethealth studies.
We describe users using features built on their activity patterns and we use off-the-shelve regressors to predict individual Big5 personality traits~\cite{Goldberg1990}.
The ground truth is the Big Five Inventory~\cite{John2008}, a personality assessment survey filled in by each student.
We compare our method with two baselines.
The first baseline~\cite{Monsted2018} uses a wide range of complex features (including autoregressive) based on call history.
The second baseline uses 145 features (\Netsense) and 2212 features (\Nethealth) self-reported in the surveys: grades, health, happiness, activity, book reading, and club membership.
These are highly personal and sensitive features, and have been shown to be the upper bound of automatic personality traits prediction~\cite{Novikov2021}.
We find our method to outperform the first baseline and, surprisingly, very competitive with respect to the prediction upper-bound.
This indicates that the Hawkes-modelled call activities embed a surprising amount of personal information.







\textbf{The main contributions of this paper are as follows:}
\begin{itemize}
	\item We show that a Hawkes point process with a power-law decaying kernel can model the phone call contact series between individuals.
	\item We use the fitted parameters of a Hawkes model to distinguish between types of relations (such as family, friendship, or romantic).
	\item We show that the call activity (modeled using Hawkes point processes) might be as predictive of the user's Big5 psychometric traits as the user filled-in questionnaire.
\end{itemize}

\textbf{The ethics of personality profiling.}
Personality profiling---particularly social media-inferred personality traits---is sometimes seen as Pandora's box.
On the one hand, personality dispositions are associated with happiness, physical and psychological health, the quality of relationships
with peers, family, and romantic others, as well as 
and community involvement~\cite{Ozer2006}, criminal activity, and political ideology at a social, institutional level~\cite{Ozer2006}.
Personality traits are also predictive of three critical life outcomes: mortality, divorce, and occupational attainment~\cite{Roberts2007}.
Such research shows the positive aspects of personality profiling research: one could build systems to prevent and improve individuals' mental health issues or their relationship with the community.
On the other hand, it was also shown that persuasive messaging is more effective when tailored for individuals' psychological characteristics~\cite{Matz2017} and that the same processes used to infer personality traits from social data can leak sensitive information such as ethnicity, political and sexual orientation~\cite{Kosinski2013}.
While social media privacy settings could (at least theoretically) bring some of the data back under the user's control, our research shows that personality traits can be inferred from data sources entirely outside the user's control.
This work adds to the understanding of what can be achieved with call logs data and advocates creating policy regulating its usage.
The latter is increasingly important, as more and more calls are being made using outside the traditional communication networks and onto online messaging platforms such as WhatsApp, which are currently not nationally regulated.


\hypertarget{sec:prerequisites-hawkes-processes}{\section{Prerequisites: Hawkes Processes}}
\label{sec:prereq}


In this section, we briefly review the theoretical prerequisites concerning modeling event series using Hawkes point processes.

\noindent{\bf Event series and point processes.}
An event is a tuple (timestamp, event features), where the \emph{timestamp} sits along the non-negative time axis, and the \emph{event features} are any descriptors related to the event.
For example, an event can be the reception of a phone call~\cite{takaguchi2012importance, raeder2011predictors}, and the features could be the caller id, length of the call, or whether it was answered or not.
An event series is a sequence of events $t_1, t_2, \ldots$, where $t_i$ are the event timestamps of the of the $i^{\text{th}}$ event, relative to the first event ($t_0 = 0$).
For ease of notation, in this paper, we use $t_i$ to denote both the event timestamp and the event itself.
A point process is a random process whose realizations consist of event series.
We denote an event series observed up to time $T$ as $\His(T) = \{t_0, t_1, \dots\}$.
%

\noindent\textbf{The Hawkes processes.} 
Hawkes processes~\cite{hawkes1971spectra} are a type of point process with the self-exciting property, i.e., the occurrence of past events increases the likelihood of future events. 
This results in the cluster property of the Hawkes property~\cite{Hawkes1974}, which states that events modeled by Hawkes appear to be grouped in time.
This latter property makes Hawkes processes desirable to model human interaction activity, which is known to follow a bursty pattern~\cite{Zipkin2016}.
An alternative approach for modeling interactions are Wold point processes defined through a Markovian transition probability distribution on the inter-event times~\cite{etesami2021variational}.
The occurrence of events in a Hawkes process is controlled by the event intensity function:
\begin{equation}
    \lambda(t \mid \His(T)) = \mu(t) + \sum_{t_i < t} \phi(t - t_i)
\end{equation}
where $\mu(t)$ is the background intensity function and $\phi: \Real^+ \rightarrow \Real^+$ is a kernel function capturing the decaying influence of a past event.
$\mu(t)$ is the immigrant intensity (i.e. the volumes of events that come from outside the system).
Here, it is constant for each call series, i.e., $\mu(t)= \mu$.
Two widely adopted parametric forms for the kernel function $\phi$ include the exponential function $\phi_{EXP}(t) = \kappa \theta e^{-\theta t}$ and the power-law function $\phi_{PL}(t) = \kappa (t + c)^{-(1+\theta)}$.

\noindent\textbf{The branching factor} $n^*$ is defined as the expected number of events directly spawned by a single event, i.e., $n^* = \int_0^{\infty} \phi(\tau) d\tau$.
$n^*$ defines the \emph{regime} of the Hawkes process:
when $n^* > 1$, each event generates more than one event; 
the process is expected to generate an infinite number of events.
When $n^* < 1$, the process generates a finite number of events, and it is expected to die out.

\noindent\textbf{Parameter estimation.} 
We estimate the parameters of the Hawkes process by maximizing the log-likelihood function for point processes~\cite{Daley2008}.
In a nutshell, the function quantifies the probability that the model generated the observed data.
Higher probabilities indicate better fit models.
Formally, 
\begin{equation} \label{eq:ll}
    \mathcal{L}(\Theta \mid \His(T)) = \sum_{t_j \in \His(T)} \log \lambda(t_j \mid \His(T)) - \int_{0}^T \lambda(\tau \mid \His(T)) d\tau
\end{equation}

\noindent{\bf Joint modeling of event series.} 
When analyzing the event series relating to a single entity (say, all the phone call series generated by the same individual), it is desirable to account for the multiple event series simultaneously.
Kong et al~\cite{Kong2020a} proposed to jointly model a group of retweet cascades with a shared Hawkes process model by summing the log-likelihood functions of individual series. 
In \hyperlink{sec:inferring-psychometric-traits}{Section \textit{Inferring Psychometric Traits}}, we jointly model the call sequences initiated by the same user, and we link the learned models to users' psychometric traits.

%


\hypertarget{sec:data-and-methods}{\section{Data and Methods}}

In this section, we first introduce \Netsense and \Nethealth, the two datasets used in this work (\hyperlink{sec:datasets}{Section \textit{Datasets}}).
Next, we build the relationship labels (\hyperlink{sec:relationships}{Section \textit{Relationships}}),
we profile the datasets (\hyperlink{sec:datasets-profiling}{Section \textit{Datasets profiling}}),
and we fit Hawkes processes to telecommunication data (\hyperlink{sec:analyzing-call-patterns-using-hawkes}{Section \textit{Analyzing call patterns using Hawkes}}).

\hypertarget{sec:datasets}{\subsection{Datasets}}
\label{subsec:dataset}

This research uses two datasets: \Netsense~\cite{striegel2013lessons} and \Nethealth~\cite{liu2018network}. The \Netsense project lasted for two and half years and gathered metadata regarding phone calls, demographic, and networking information about college students enrolled in the study during 2011--2013.
The \Netsense study was followed by \Nethealth, which also recorded metadata on mobile phone activities of Notre Dame University students but over a longer time -- four and a half years from 2015 to 2019.
Both datasets consist of two parts: the calling and texting activity and the student surveys.

\textbf{Ethics}. All methods were carried out following relevant guidelines and regulations. The Institutional Review Board (IRB) of the University of Notre Dame has reviewed the \Netsense study and has approved it - the IRB Number is FWA 00002462. 
The observational study \Nethealth was also approved by the University of Notre Dame's IRB after a full board review under protocol 17-05-3912. 
All participants provided written informed consent prior to taking part in both studies.


\textbf{Calling and texting activity.}
The \Netsense and \Nethealth datasets record all calls and texts from the students' phones enrolled in the study, both outgoing and incoming.
The dataset records the caller id, the receiver id, the timestamp, and the call duration for each phone call.
For text messages, it records the sender, the receiver, and the timestamp.
\Nethealth also records metadata about WhatsApp and iMessage communication.
Where both the sender (or caller) and the receiver are students enrolled in the program, the call (or text) will be recorded twice -- once for the caller and once for the receiver.
\Netsense and \Nethealth also record calls and texts from people outside the study (such as family and friends).

\textbf{The surveys.}
The datasets include detailed data on the students since each participating student was surveyed every term about their interests, opinions, or relationships with others.
The students have been surveyed six times for \Netsense and eight times for \Nethealth.
This paper leverages two aspects of these surveys: how students describe their relationships with others and how they represent themselves.
For the former, participants labeled the relationship with their peers using descriptors such as a friend, family, significant other, co-worker, other, or not labeled at all.
For the latter, the students provided information about their hobbies, activities, well-being, grades, weight and height, health condition, and the number of books they read.
We use this information in \hyperlink{sec:inferring-psychometric-traits}{Section \textit{Inferring Psychometric Traits}} to build a baseline for predicting personality traits.

\textbf{The Big5 psychometric questionnaire.}
For both datasets and with each survey, students provided answers to forty-four questions from the Big Five Inventory~\cite{John2008}.
Students answered the questions on whether they would describe themselves as \textit{talkative}, \textit{curious about many different things}, \textit{tending to be quiet}, among others.
Their answers can map each student as a point in the big five personality traits space~\cite{Goldberg1990}, where each traits (openness, conscientiousness, extraversion, agreeableness, neuroticism) is represented as a numeric value between one and five.

%

%
\begin{table}[tbp]
  \small
  \caption{
    The rules for categorizing relationships used in this study and the number of obtained instances for each rule.
  }
  \label{tab:relationships-categorization}
  \setlength{\tabcolsep}{1.5pt}
  \begin{tabular}{llp{1cm}rr}
    \toprule
    Previous relation & New relation & Label                                     & \Netsense & \Nethealth \\
    \midrule
    friend            & sibling      & \multirow{3}{*}{\makecell[cl]{family-                              \\relaxing}} & \multirow{3}{*}{115} & \multirow{3}{*}{313} \\
    friend            & parent       &                                           &                        \\
    friend            & other family &                                           &                        \\ \midrule
    parent            & parent       & \multirow{2}{*}{\makecell[cl]{family-                              \\stable}} & \multirow{2}{*}{363} & \multirow{2}{*}{422} \\
    sibling           & sibling      &                                                                    \\ \midrule
    friend            & coworker     & \multirow{3}{*}{\makecell[cl]{friendship-                          \\relaxing}} & \multirow{3}{*}{569} & \multirow{3}{*}{1709} \\
    friend            & other        &                                                                    \\
    friend            & acquaintance &                                                                    \\ \midrule
    friend            & friend       & friendship-stable                         & 675       & 988        \\ \midrule
    other             & friend       & \multirow{3}{*}{\makecell[cl]{friendship-                          \\strengthening}} & \multirow{3}{*}{41} & \multirow{3}{*}{74} \\
    coworker          & friend       &                                                                    \\
    acquaintance      & friend       &                                                                    \\ \midrule
    significant other & other        & \multirow{2}{*}{\makecell[cl]{romantic-                            \\relaxing}} & \multirow{2}{*}{16} & \multirow{2}{*}{18} \\
    significant other & friend       &                                                                    \\
    \bottomrule
  \end{tabular}
\end{table}

\begin{figure*}[tbp]
	\centering
	\setkeys{Gin}{height=0.14\textheight}
	\subfloat[]{
		\includegraphics{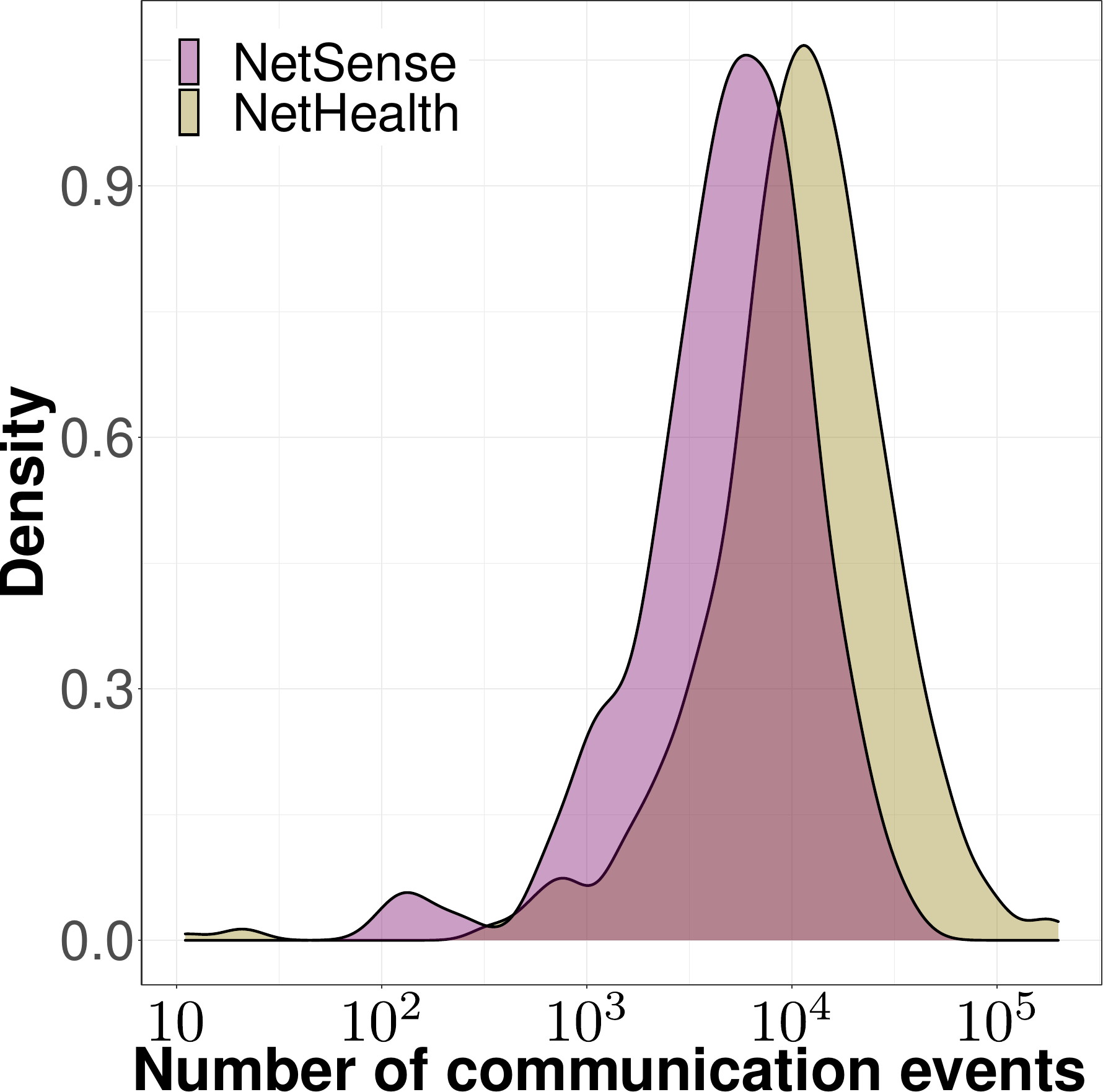}
		\label{fig:profiling1}
	}%
	\subfloat[]{
		\includegraphics{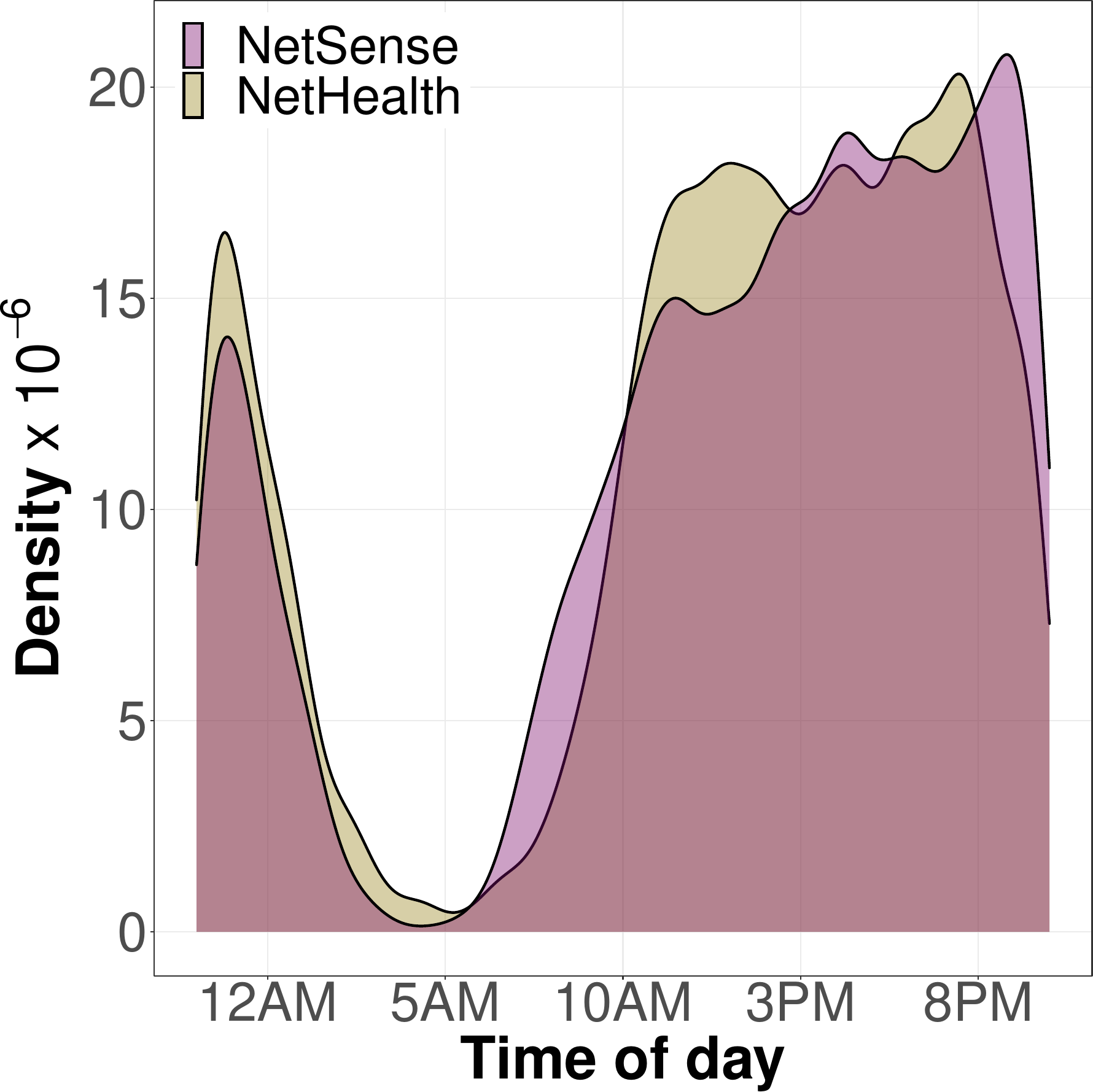}
		\label{fig:profiling2}
	}%
	\subfloat[]{
		\includegraphics{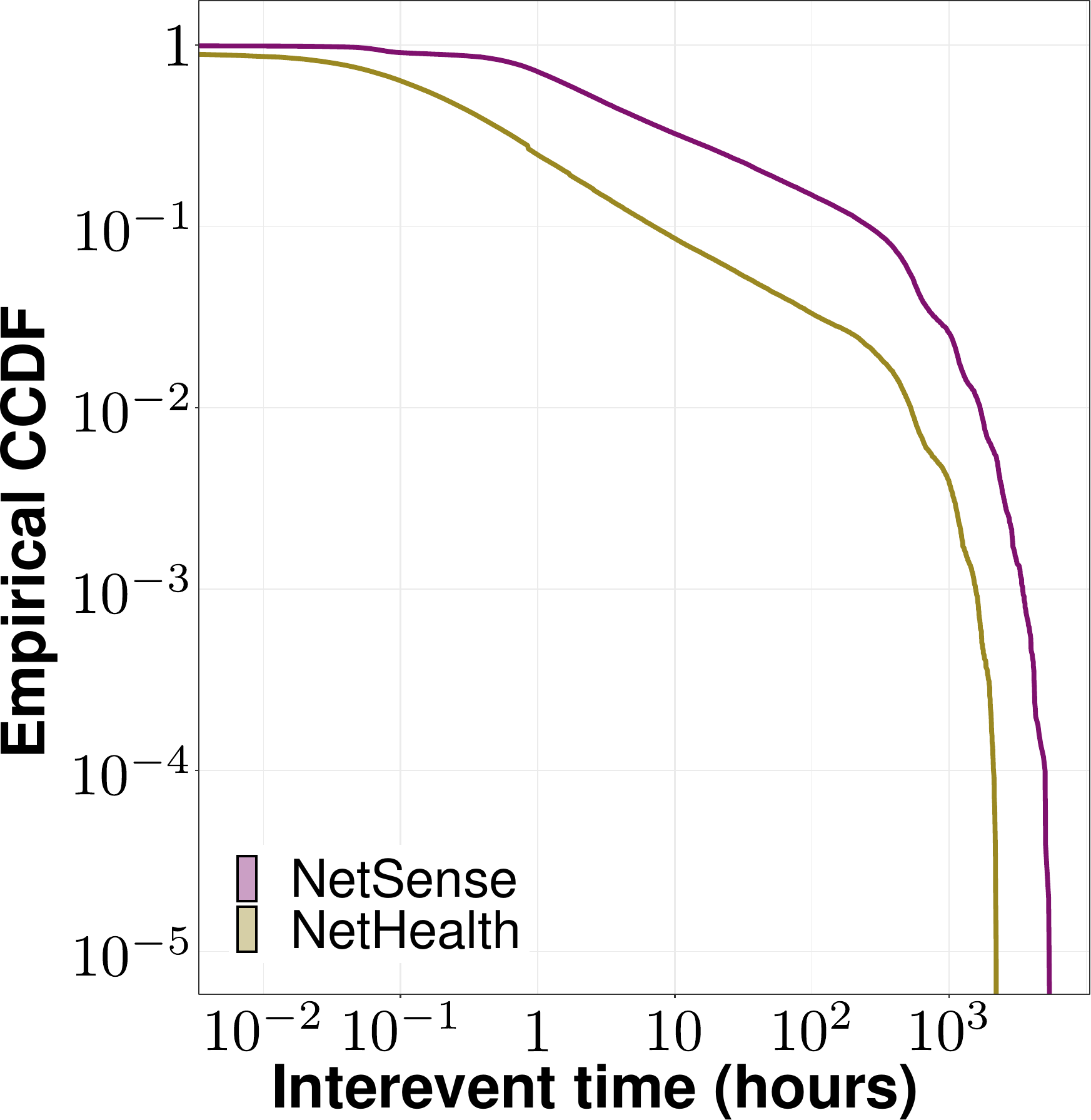}
		\label{fig:profiling3}
	}
	\subfloat[]{
		\includegraphics{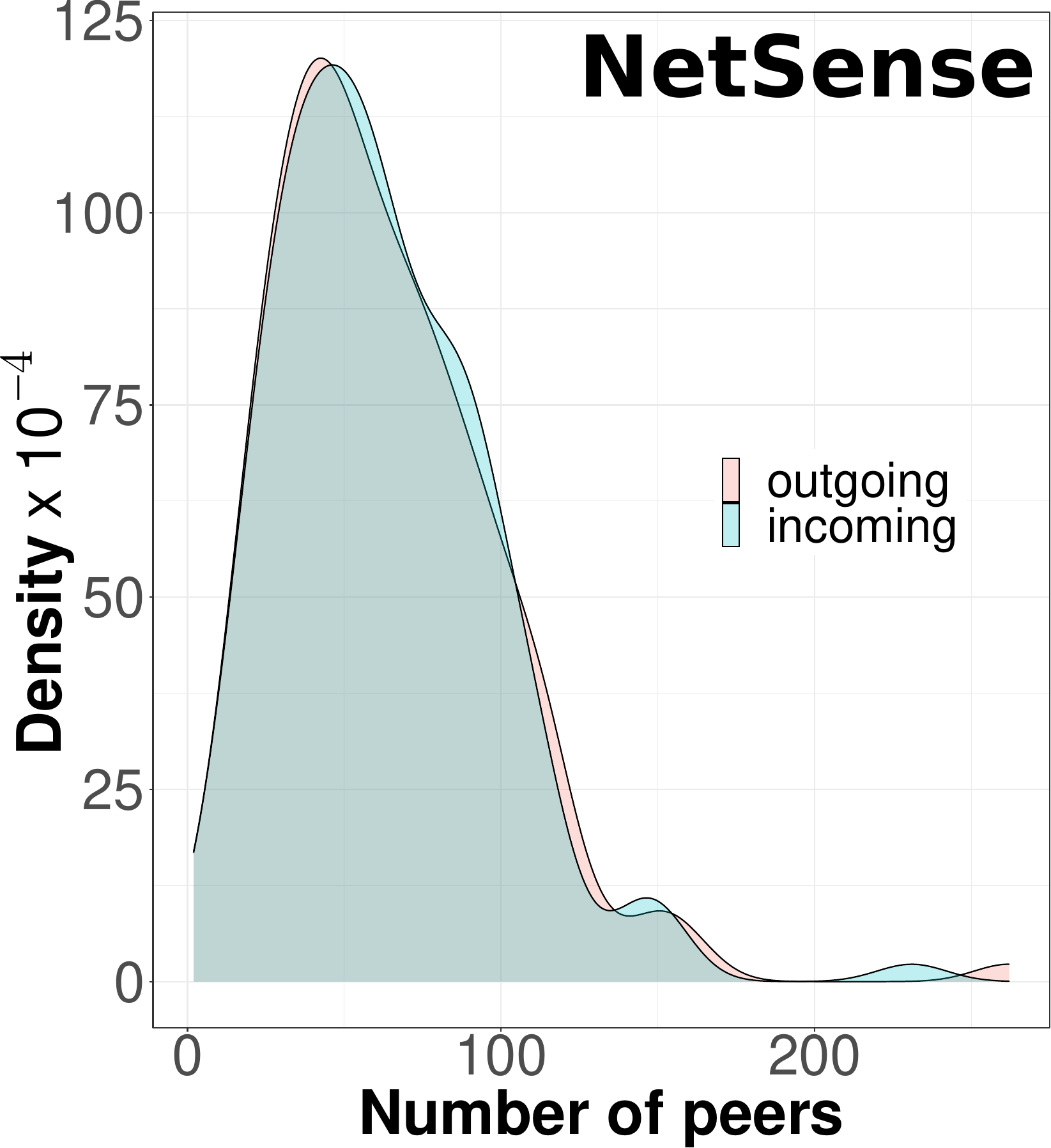}
		\label{fig:profiling4}
	}
	\subfloat[]{
		\includegraphics{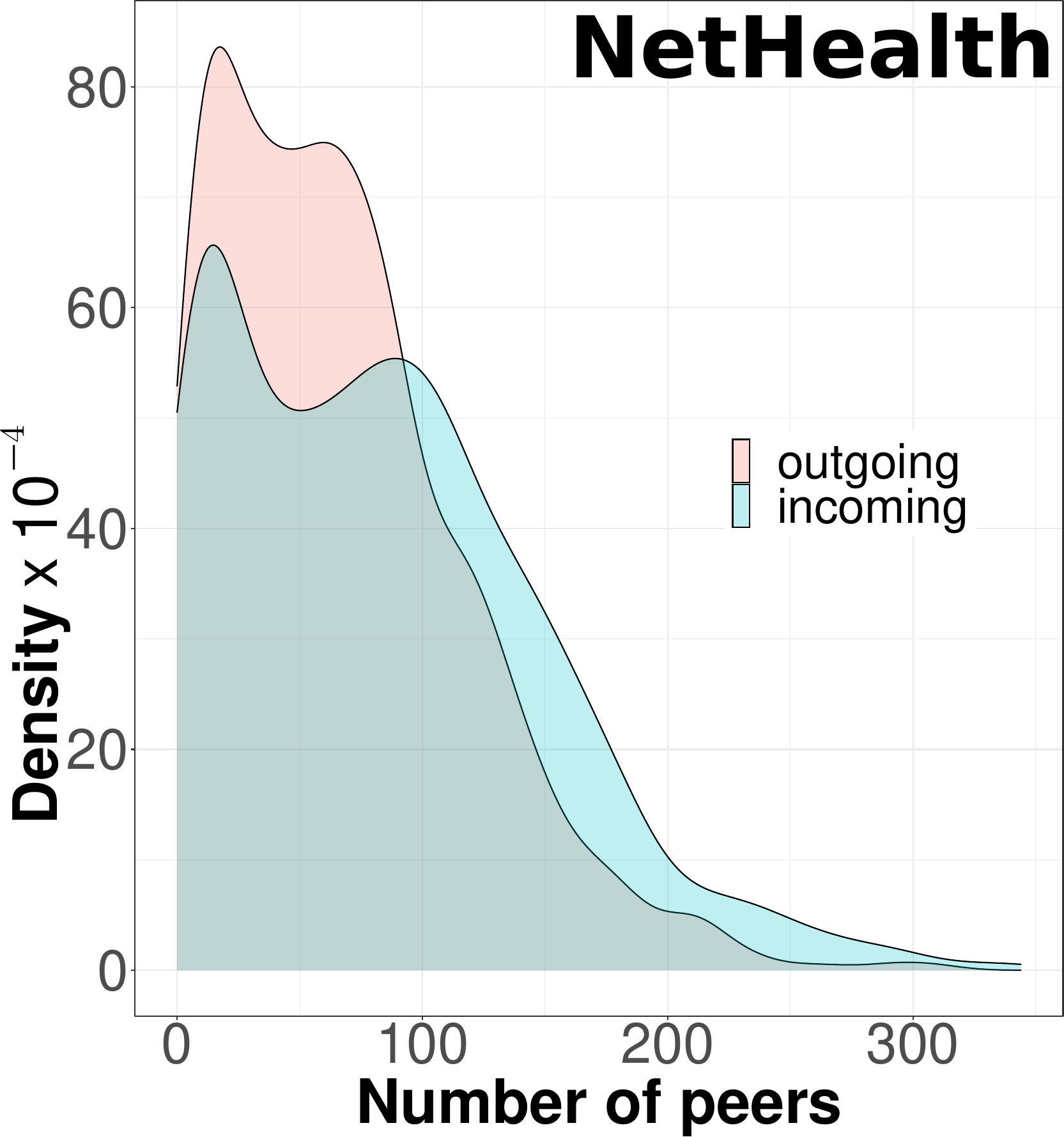}
		\label{fig:profiling5}
	}
	\caption{
		Exploratory analysis of the communication data for students participating in the \Netsense and \Nethealth studies at the University of Notre Dame.
		(a) density plot for the number of communication events for the students, 
		(b) the distribution of phone activity through the course of the day, 
		(c) the cumulative distribution function of the inter-event times in hours, 
		(d) and (e) the number of incoming and outgoing communication peers for the students in \Netsense and \Nethealth studies, respectively, where there are at least twenty communication events with the peer during the analyzed period.}
	\label{fig:profiling}
\end{figure*}

\hypertarget{sec:relationships}{\subsection{Relationships}}
\label{sec:relationships}
Research in Sociology, Psychology, Communication, and related fields show that social network relations are inherently dynamic -- they change in strength, quality, and character over time \cite{Suitor1997-cq,Suitor1997-jl,Feld2007-gk,Rivera2010-fg,gilbert2009predicting}. 
Strong ties can become weak \cite{Burt2000-kl}, and weak ties can become strong \cite{flamino2021machine}. Contacts may transition from initial acquaintances or workmates to being close friends \cite{Feld2007-gk, Wellman1997-pg}. 
In the same way, geographical separation can change the character of ties, with close friends devolving into casual ties. 
The fundamental message of this previous work is that people's relations with their peers evolve.

To capture these dynamics, the \Netsense and \Nethealth projects queried students about their personal relationships with their top-twenty contacts six times during the respective study periods. 
In particular, each participant was asked to provide a relationship \emph{label}, characterizing each tie as a friend, family (e.g., parent, sibling, other kin), co-worker, acquaintance, or significant other.
%
We discriminate between three types of temporal dynamics for relationships: stable, strengthening, and relaxing.
We denote a relation as stable when the student labels it identically across all surveys.
When the relationship transitions between categories, we annotate it with the type and direction of the development, following the rules listed in \cref{tab:relationships-categorization}.
We consider a \emph{friend} as a stronger type of relation than a \emph{acquaintance};
therefore we label a transition \emph{acquaintance} $\rightarrow$ \emph{friend} as ``friendship-strengthening''.
Similarly, we denote the transition \emph{significant other} $\rightarrow$ \emph{friend} as ``romantic-relaxing''.
One type of transition that might appear peculiar is the ``family-relaxing'', where the relation is initially labeled as \emph{friend}, and later on, it becomes \emph{sibling}, \emph{parent} or \emph{other family}.
The reader should remember that the subjects of the studies were college students, for most of whom college is the first time they leave home and live independently.
Therefore, it is not inconceivable that the relationship with the family members can go colder over time as they discover friends and college life.
Finally, we discard the 13 relationships which have more than one type transition; we consider these as likely input errors by the students (e.g., \textit{friend} $\rightarrow$ \textit{significant other} $\rightarrow$ \textit{family} $\rightarrow$ \textit{other}).
We also discard the relationship classes with less than 15 instances to have enough instances in each class for building classifiers in \hyperlink{sec:classify-relationship-types}{Section \textit{Classify relationship types}}.
\cref{tab:relationships-categorization} shows each relationship type's final volumes in each dataset.


\hypertarget{sec:datasets-profiling}{\subsection{Datasets profiling}}
\label{sec:profiling}

The \Netsense dataset covers 178 students who filled out the surveys and whose mobile phone communication is available.
\Netsense records 7,575,864 phone interactions using the application installed on the students' mobile phones.
The \Nethealth study covered 712 students, and it collected information about 60,486,565 phone interactions.
In \cref{fig:profiling}, we present the exploratory analysis of communication datasets.
\cref{fig:profiling1} shows the density of the number of communication events per student. Visibly, the distributions look very similar for both datasets, with most students recording between 5,000 and 20,000 communication events.
The distribution is shifted towards higher values for \Nethealth as this study lasted longer than \Netsense.
\cref{fig:profiling2} shows the density of communication events with respect to the time of day when they were initiated.
Most communication starts around noon and continues up to 11 PM.
\cref{fig:profiling3} plots the empirical cumulative distribution of the inter-event times -- we observe that these inter-event times are long-tailed distributed, a result already known in literature~\cite{Jo2012}.
Lastly, \cref{fig:profiling4} and \cref{fig:profiling5} present for \Netsense and \Nethealth, respectively, the number of communication peers for each of the students participating in the survey,
We distinguish between incoming and outgoing peers and filter out the peers with less than 20 interactions to remove occasional phone calls.
Visibly, for \Netsense, the two density plots mostly overlap; however, for \Nethealth, the students appear to have more incoming peers -- probably reflecting the changes in how college students communicate as the conversations involved many short messages on platforms like iMessage.
For both studies, most students have less than 130 peers, which is compatible with Dunbar's number and the theory on cognitive limits of the human brain~\cite{dunbar1992neocortex}.
For \Netsense, out of the 3,159,669 total outgoing communication events, 2,012,816 have been interactions with peers that students labeled in surveys with the type of the relationship.
That gives 1,779 relationships to investigate for \Netsense.
For \Nethealth, out of 23,705,245 outgoing events, 15,605,131 were with 3,524 peers that have been categorized -- the number of instances for each category is shown in \cref{tab:relationships-categorization}.
The following section fits a Hawkes model for each communication relationship.



%
\begin{figure*}[tbp]
	\centering
	\setkeys{Gin}{height=0.205\textheight}
	\subfloat[]{
		\includegraphics{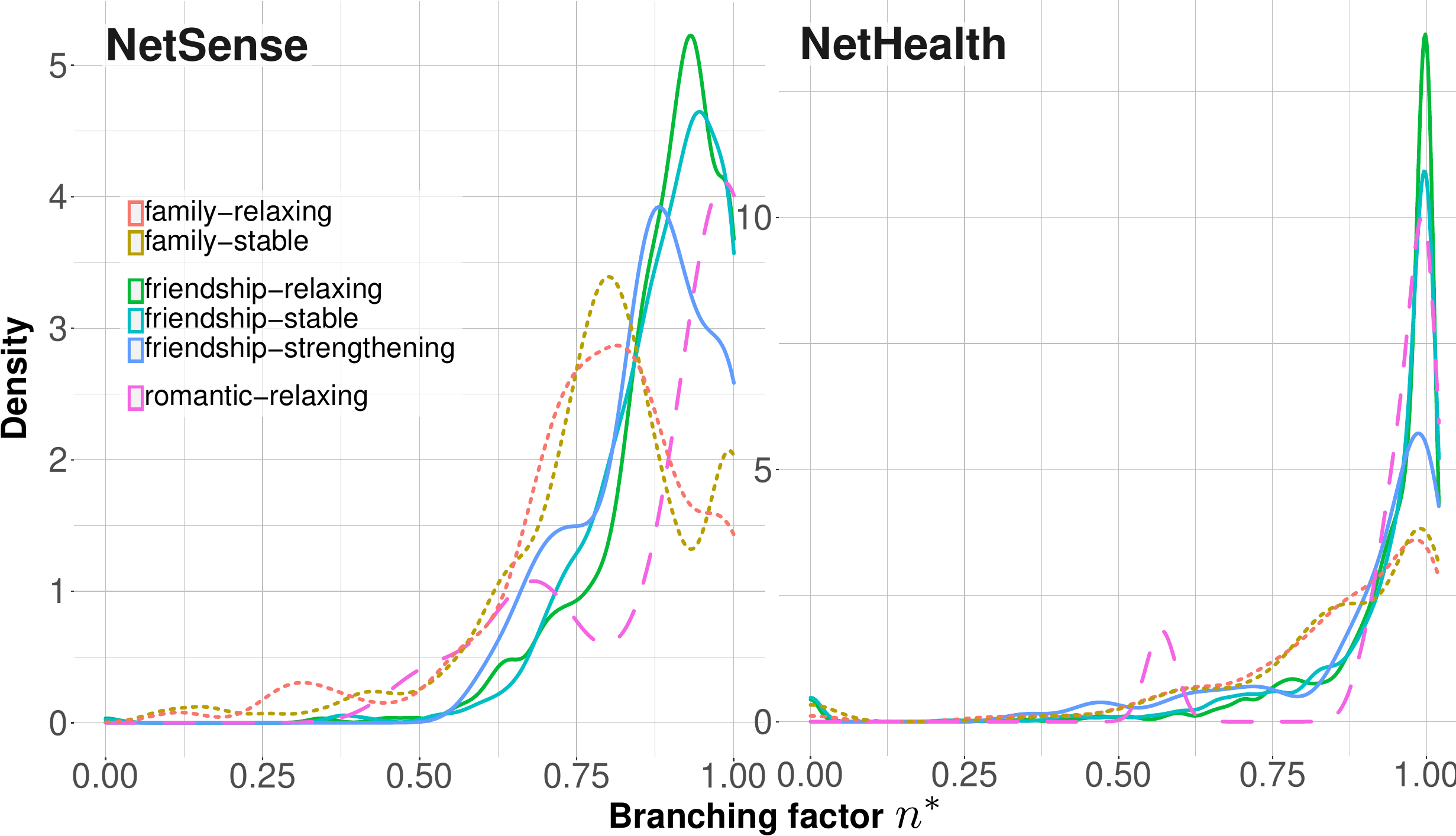}
		\label{fig:paramBranchingFactor}
	}%
	\subfloat[]{
		\includegraphics{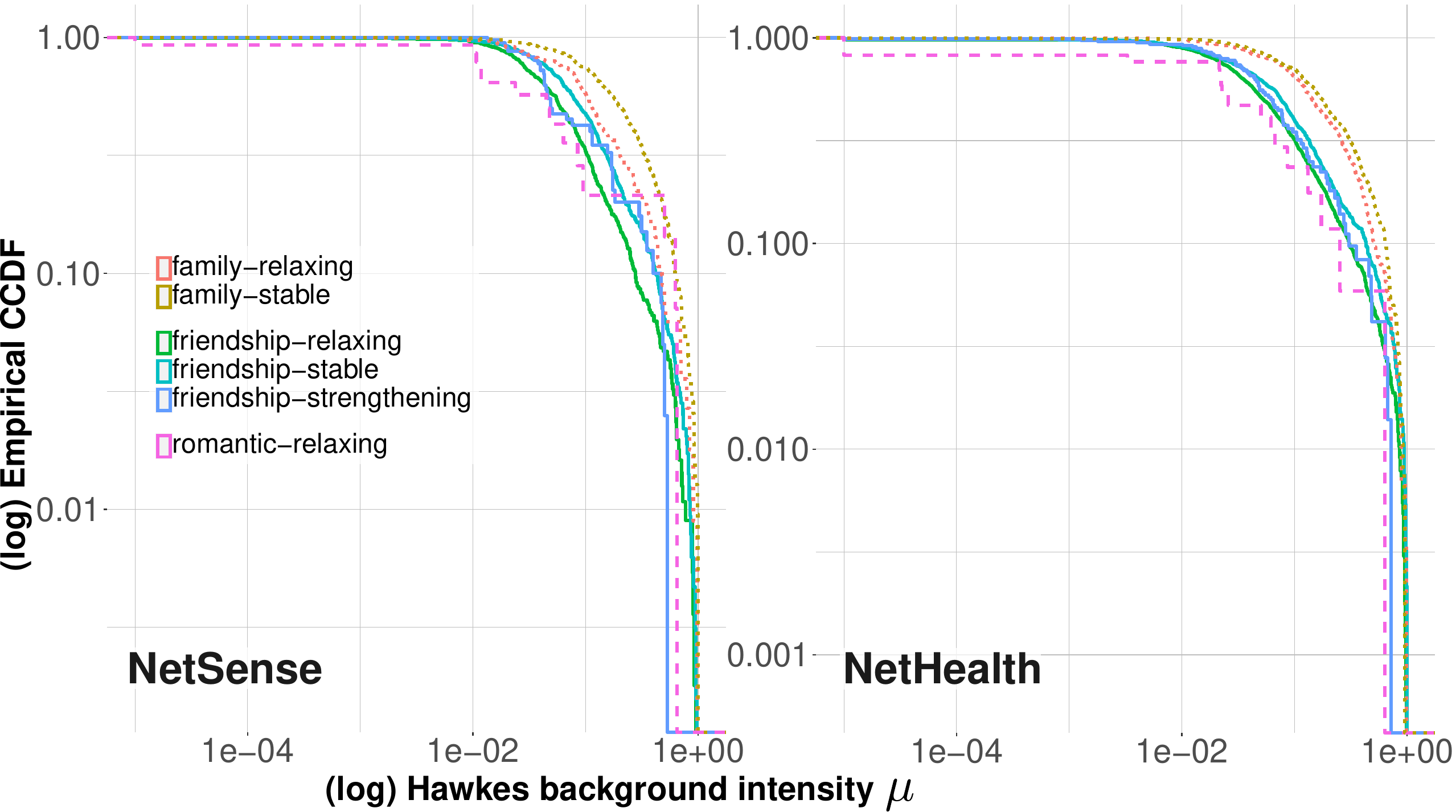}
		\label{fig:paramMu}
	}%
	\caption{
		Distribution of Hawkes process parameters per relationship types:
		(a) the density plot for the branching factor $n^*$, 
		(b) the complementary cumulative distribution function of the background event intensity parameter $\mu$.
		Family relations are shown in dotted lines, friendship in solid lines, and romantic in dashed lines.
	}
	\label{fig:hawkesParams}
\end{figure*}

\hypertarget{sec:analyzing-call-patterns-using-hawkes}{\subsection{Analyze call patterns using Hawkes}}

\label{sec:fitting}
We model the call series between two individuals as a point process.
We assume that future calls are more likely when other calls have recently occurred -- in other words, the calling process has the self-exciting property.
This hypothesis implies that the future of a sequence of events depends on its entire past and not only the current state.
In other words, it does not have the Markov property.
Consequently, typical choice modeling tools (such as Markov chains models) are unsuitable for modeling calls.
Instead, we fit the call series using Hawkes point processes -- which is a well-understood non-Markovian extension of the Poisson process~\cite{Rizoiu2022}.
We leverage a general-purpose point process \texttt{R} library (\texttt{evently}~\cite{Kong2020b}) to produce user-level descriptions based solely on the fitted model parameters of the call series.

\noindent\textbf{Map calls to point processes.}
In this work, we use phone calls and texts (both denoted hereafter as calls) as \emph{point process events} (defined in \hyperlink{sec:prerequisites-hawkes-processes}{Section \textit{Prerequisites: Hawkes Processes}}).
Note that \Nethealth includes the WhatsApp and iMessage group messages, which we remove since they are not peer-to-peer communication and are likely to follow other dynamics.
An event in our datasets is a tuple (timestamp, sender, receiver).
We build \emph{event series} by grouping incoming and outgoing events between pairs of users.
Each timestamp $t_i$ is the time difference (in days) between the recorded timestamp of calls and the timestamp of the first call ($t_0 = 0$).
We retain only the series where the sender is a student enrolled in the study.
We impose the latter condition so that, in \hyperlink{sec:inferring-psychometric-traits}{Section \textit{Inferring Psychometric Traits}}, we can match the call series with the surveys filled in by the students.
When both the sender and receiver are students in the study, the outgoing series for the sender is identical to the incoming series for the receiver.
We filter our call series with fewer than 20 or more than 10,000 recorded calls or texts.
We require the former condition to have enough data to fit the Hawkes processes (described next).
The latter condition is to avoid computational explosion (considering that Hawkes is quadratic with the number of events).
This results in 15,110 call series associated with 178 students, totaling more than 5 million call events for \Netsense.
For \Nethealth, we get 70,752 call series associated with 712 students, totaling more than 33 million events.

\noindent\textbf{Fit Hawkes processes.}
For each obtained call series, we fit the parameters of a Hawkes model, using the exponential and the power-law kernel functions ($\phi_{EXP}(t)$ and $\phi_{PL}(t)$, respectively, defined in \hyperlink{sec:prerequisites-hawkes-processes}{Section \textit{Prerequisites: Hawkes Processes}}).
We consider the exogenous intensity function to be constant ($\mu(t) = \mu$), and we fit it from data together with the other parameters using the software package \texttt{evently}~\cite{Kong2020b}.
\texttt{Evently} is a \texttt{R} package for modeling events series using Hawkes processes and their variants.
Internally, \texttt{evently} leverages \texttt{IPOPT}~\cite{Wachter2006} -- the current state of the art in constrained, non-linear optimization -- to maximize the log-likelihood function in \cref{eq:ll}.
By design, it supports a wide array of kernel functions and provides an integrated set of functionalities to conduct event series-level or aggregated-level analyses.
For each event series, \texttt{evently} outputs the fitted parameters $\mu$, $\kappa$, $\theta$ (and $c$ for $\phi_{PL}(t)$), and the branching factor $n^*$.

\noindent\textbf{Interpreting the Hawkes parameters.}
The Hawkes process is a generative model with interpretable parameters.
The branching factor is a crucial quantity for point processes. 
It indicates the size of the self-exciting effect -- when $n^*$ is large, previous calls generate more future calls, and the recent calling history mainly drives the calling relation.
The background intensity indicates the size of the exogenous effect -- when $\mu$ is large, calls are driven by outside factors.
For example, one would expect that the call series between parents and children would exhibit a high $\mu$, e.g., by calling daily at the same time.
The parameter $\theta$ controls the decay rate of the call generation -- when $\theta$ is small, the kernel function $\phi(t)$ decays slowly, and new calls are generated endogenously for more extended periods.

\noindent\textbf{Build user representation.}
We construct user descriptions using the methodology described by Kong et al~\cite{Kong2020}.
We jointly fit all call series involving a given user and describe each user's activity with a latent mixture of Hawkes models.
Due to Hawkes' quadratic complexity, the joint fitting has significant computation speed advantages over a single large Hawkes model containing all user events.
We build the user description by first computing the 5\% percentiles bins for every parameter (i.e., $\mu$, $\kappa$, $c$, and $\theta$) for all mixture components and all users.
Next, we compute the percentage of mixture parameters within each bin for each user.
This results in 20 values (one per 5\% percentile bin) for each parameter.
In the end, we add a summary of temporal features for user series -- the six-point summaries (min, mean, median, max, $25^{th}$ and $75^{th}$ percentile) of the call inter-arrival times, the total number of calls in a series and the duration of a series (in days).
We also add the total number of call series in which a user is involved.
This results in a 99-length vector describing user calling activity.








%
\begin{table}[tbp]
    \caption{
        Percentage of call series (for each dataset) passing the goodness of fit tests (p-value > 0.05, see details) and median holdout log-likelihood per event (LL) for the two Hawkes kernels. Higher is better.
        The winner kernel is shown in bold.
    }
    \label{tab:goodness-of-fit}
    \setlength{\tabcolsep}{2.5pt}
    \begin{tabular}{ll|rrr|r|r}
        \toprule
                                    &             & \multicolumn{1}{c}{KS} & \multicolumn{1}{c}{ED} & \multicolumn{1}{c|}{LB} & \multicolumn{1}{c|}{KS wins} &
        \multicolumn{1}{c}{LL}                                                                                                                                                \\ \midrule
        \multirow{2}{*}{\Netsense}  & Exponential & 50.38\%                & 37.30\%                & 88.62\%                 & 20.90\%                      & 0.401          \\
                                    & Power-Law   & \textbf{68.12\%}       & \textbf{58.63\%}       & 82.49\%                 & \textbf{78.10\%}             & \textbf{0.665} \\ \midrule
        \multirow{2}{*}{\Nethealth} & Exponential & 39.22\%                & 31.03\%                & 89.60\%                 & 22.50\%                      & 0.366          \\
                                    & Power-Law   & \textbf{61.74\%}       & \textbf{58.30\%}       & 87.59\%                 & \textbf{74.50\%}             & \textbf{0.666} \\
        \bottomrule
    \end{tabular}
    \label{}
\end{table}

\hypertarget{sec:select-a-hawkes-process-kernel}{\section{Select a Hawkes process kernel}}
\label{subseq:fittingHawkes}

Here, we select the Hawkes kernel that best describes the time-decaying influence of calls in telecommunication series.
We use goodness-of-fit tests and generalization error to compare the two kernels commonly used with human traces data -- $\phi_{EXP}(t)$ the exponential and $\phi_{PL}(t)$ the power-law.

\noindent\textbf{Goodness-of-fit tests.}
For point processes, the random time change theorem~\cite{brown2002time} states that the inter-event times transformed using the compensator (i.e., the definite integral of the event intensity) should follow a unit rate exponential distribution (see \cite{Laub2015,Kong2020} for more details).
We perform three \textit{portmanteau} statistical tests: the Kolmogorov-Smirnov (KS), Excess Dispersion (ED), and Ljung-Box (LB). 
For all three tests, to reject the null hypothesis (p-value < 0.05) means failing the test (that is, the sequence does not have the required property).
The KS and ED tests assess whether the transformed interevent times are drawn from a unit exponential distribution, whereas LB tests their independence.
We also compare the KS statistic for the exponential and power-law kernels to decide which fits better for the passing KS tests.

\noindent\textbf{Generalization error.}
We test the generalization of each kernel in a temporal holdout setup.
We temporally split each call series into two parts.
First, the earlier $80\%$ of the call events in each series are used to train model parameters.
Next, we compute the holdout log-likelihood on the later $20\%$ of the events.
Finally, we normalize the holdout log-likelihood by the number of events in the test set to account for the variable number of calls between series -- i.e., we compute the log-likelihood per event.

\noindent\textbf{Results.}
\cref{tab:goodness-of-fit} shows the percentage of call series that pass the statistical tests for each kernel and each dataset.
Visibly, power-law exhibits a higher percentage of passing the fitness tests (KS and ED) than exponential.
Both kernels show a high passing rate for the independence test (LB).
Out of the passing KS tests, power-law wins the highest percentage of pairs when comparing the KS statistic (KS wins)---$78.1\%$ for \Netsense and $75.4\%$ for \Nethealth.
\cref{tab:goodness-of-fit} also shows the median holdout log-likelihood per event for each kernel and each dataset (we show it summarized as boxplots in the online appendix~\cite{appendix}).
The power-law kernel outperforms the exponential kernel for both datasets.
Together with the goodness-of-fit tests, this leads us to conclude that the power-law describes the call series better than exponential.
Therefore, in the rest of this paper, we only present results for the power-law kernel with all call series refitted using all the available calls.
%

\hypertarget{sec:characterizing-relationships}{\section{Characterize Relationships}}

In \hyperlink{sec:describe-relations-using-hawkes}{Section \textit{Describe relations using Hawkes}}, we show that the fitted model parameters are descriptive for relationship types.
In \hyperlink{sec:classify-relationship-types}{Section \textit{Classify relationship types}}, we train classifiers to distinguish relationship types.

\hypertarget{sec:describe-relations-using-hawkes}{\subsection{Describe relations using Hawkes}}

\label{subseq:parameter-interpretation}

Here, we use the Hawkes parameters' interpretability and analyze the fitted Hawkes parameters per relationship type.
We concentrate on two quantitites: the branching factor $n^*$ (\cref{fig:paramBranchingFactor}) and the background event intensity $\mu$ (\cref{fig:paramMu}).
%
\cref{fig:hawkesParams} shows that the different relations types have specific exogenous and self-excitation patterns.
\emph{Family relations} (shown in dotted lines) have lower branching factors and higher background intensity values.
This indicates that outside factors mainly drive the calling dynamics between family members -- e.g., students might call their families every day, at the same time, to let them know they are safe.
\emph{Friendship relations} (shown in solid lines) appear to have the opposite pattern.
They are mainly driven by self-excitation (high $n^*$), and friends are less likely than family to call each other after a period of pause (low $\mu$) -- linking to the old adage: ``out of sight, out of mind''.
Finally, \emph{romantic relations} (dashed lines) have a peculiar mixed behavior.
Both $n^*$ and $\mu$ show two modes, one for lower values and one for higher values, indicating a dual nature of romantic relations -- some partners call each other because of recent calling activity;
in contrast, others use more of a scheduled approach.
These patterns are strongly consistent across both datasets and indicate that we can distinguish between types of relationships using a machine learning classifier (see next section).
\hypertarget{sec:classify-relationship-types}{\subsection{Classify relationship types}}
\label{sec:differentiating-relationships}

Here, we ask whether the relationships described using fitted Hawkes processes are identifiable one from another.

\noindent\textbf{Predictive setup.}
We describe each relationship using the fitted Hawkes parameters $\mu$, $\kappa$, $\theta$ and $c$, and its branching factor $n^*$. 
In addition, we train an ARIMA~\cite{Makridakis1997} model for each relationship and use the fitted parameters as a baseline to compare with the predictive performance of Hawkes features.
Next, we use the Hawkes, ARIMA, and concatenated Hawkes+ARIMA feature sets to train three off-the-shelve classification algorithms:
%
%
Random Forests~\cite{breiman2001random},
Support Vector Machines (SVM)~\cite{cortes1995support}, and
XGBoost~\cite{chen2016xgboost}.
%
We compute the prediction performance of each algorithm using two nested cross-validation loops. 
The outer loop uses five folds to estimate the generalization performance on unseen data. 
The inner loop tunes hyperparameters using 5-fold cross-validation and random search with 500 combinations.
We use the SMOTE algorithm to balance the size of the classes.

%
\begin{figure*}[tbp]
	\centering%
	\setkeys{Gin}{height=0.3\textheight}%
	\includegraphics{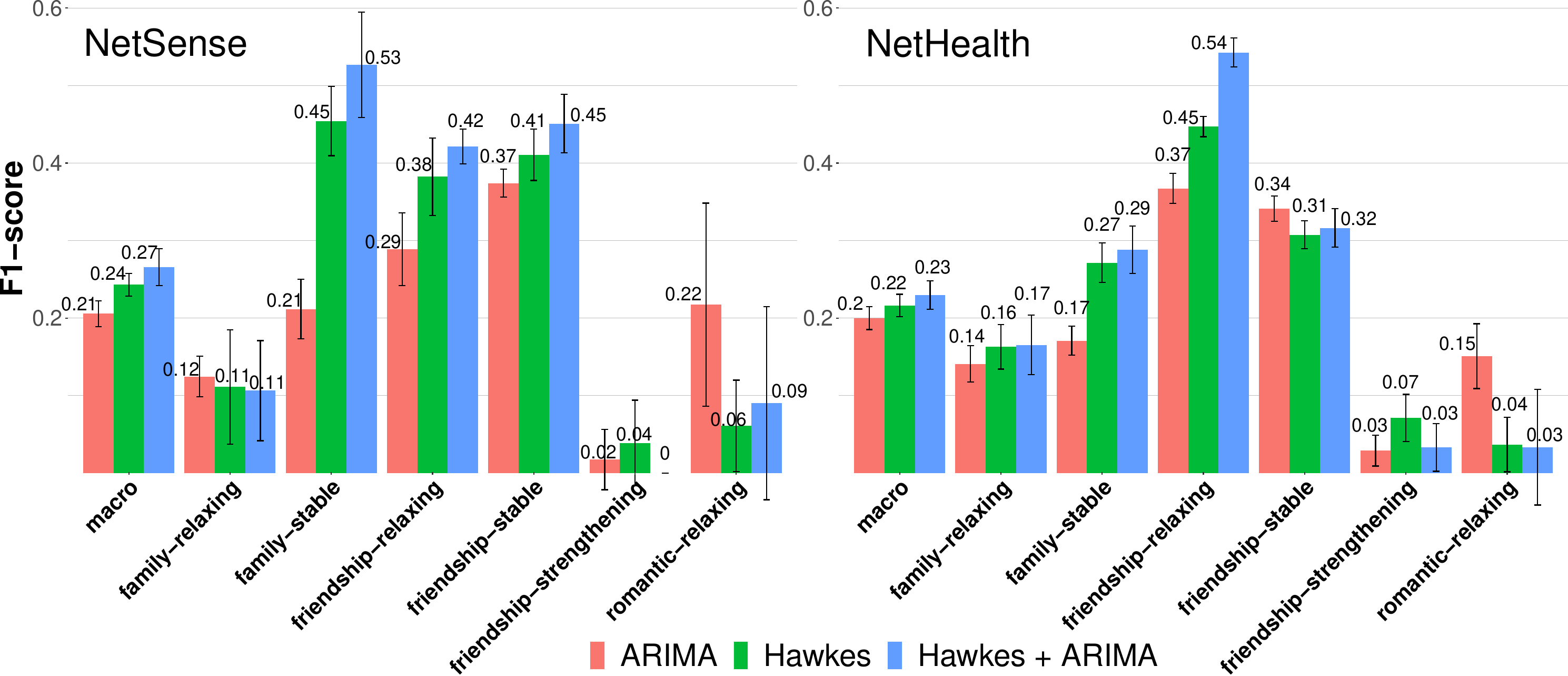}
	\caption{
		Performance of predicting relationship types using the call activity modeled using Hawkes processes and ARIMA (higher is better).
	}
	\label{fig:relationship-classification}
\end{figure*}

\noindent\textbf{Results.}
\cref{fig:relationship-classification} shows the prediction performance for Random Forest, which achieved the best result (see the results for the other two classifiers in the online appendix~\cite{appendix}).
The Hawkes features obtain a macro F1-score of $24\%$ for \Netsense and $22\%$ for \Nethealth, more than double the random classifier and outperforming the ARIMA baseline ($21\%$ and $20\%$, respectively).
Furthermore, the Hawkes and the ARIMA approaches are complementary, as the concatenated feature set Hawkes+ARIMA yields the best results.
Unsurprinsingly, the larger classes (\emph{family-stable}, \emph{friendship-stable} and \emph{friendship-relaxing}) are consistently best predicted.
It is perhaps surprising how poorly predicted is the \emph{family-relaxing} class given its size, which indicates that there are multiple dynamics of how students interact increasingly less with their family.
Overall, the results show that relationship types (and their dynamics) can be inferred from the parameter of a Hawkes process fitted on the call series. 
Additionally, we performed temporal change point detection in the dynamic relations using a paired Wilcoxon signed-rank test (conf. level = 95\%) and effect size (Cohen's d~\cite{cohn1988statistical}) to compare the log-likelihood before and after the change (more details in the online appendix~\cite{appendix}).
We find that statistically significant results (p-value $\ll 0.001$) were obtained for \Nethealth for family-relaxing (effect size = 0.25) and friendship-relaxing (effect size = 0.14).

\begin{figure*}[tbp]
	\centering
	\includegraphics[width=.99\textwidth]{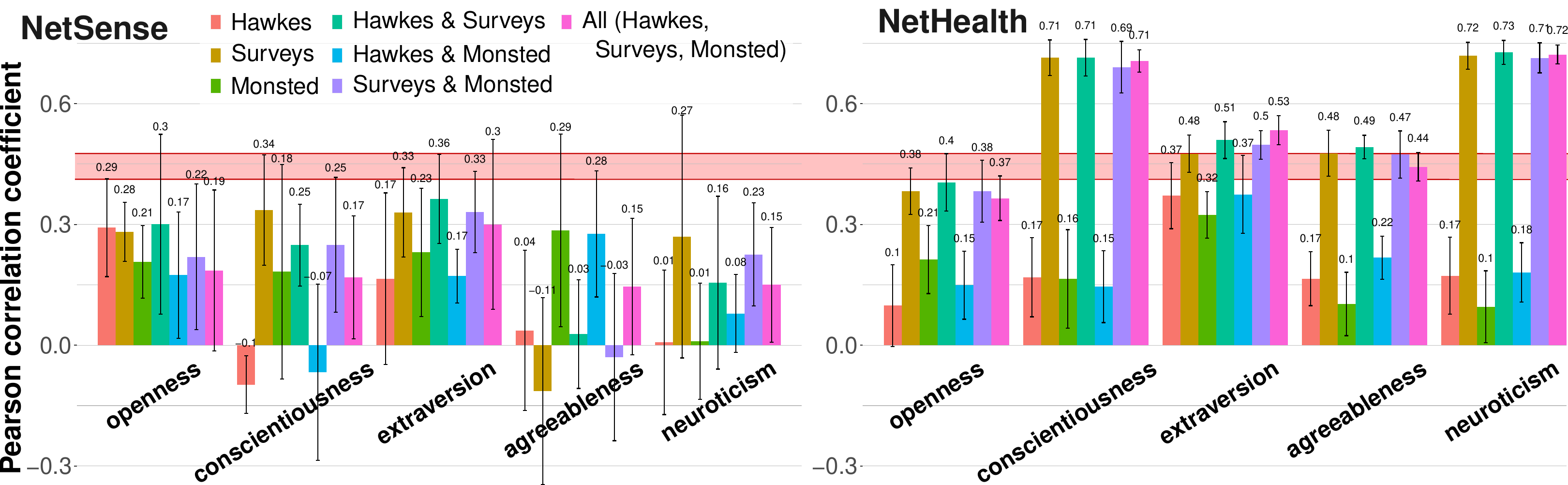}
	\caption{
		\textbf{Big5 traits prediction performance measured using the Pearson correlation coefficient.}
		Students are described using features built on two data sources two data sources (calls and filled-in surveys).
		For calls, we describe the users using our Hawkes-model embedding (Hawkes) and using a recent baseline~\cite{Monsted2018} (Monsted).
		We also test combinations of feature sets.
		The red area shows the credible upper limits for correlations between predicted and self-reported personality traits~\cite{Novikov2021}.
	}
	\label{fig:Big5-pearson}
\end{figure*}

\hypertarget{sec:inferring-psychometric-traits}{\section{Infer Psychometric Traits}}
Here, we use call activity to infer the students' Big5 profiles~\cite{Goldberg1990}.
First, in \hyperlink{sec:baselines-and-feature-sets}{Section \textit{Baselines and feature sets}} we detail the baselines and sets of features.
Next, in \hyperlink{sec:baselines-and-feature-sets}{Section \textit{Inferring psychometric traits}}, we present the prediction performance.

\hypertarget{sec:baselines-and-feature-sets}{\subsection{Baselines and feature sets}}

\label{subsec:baselines}

We set out to predict personality traits using three sets of features, which capture the different aspects of user activity.
The first set is the Hawkes-based user representation that we introduced in \hyperlink{sec:analyzing-call-patterns-using-hawkes}{Section \textit{Analyzing call patterns using Hawkes}}.
These features use the call series solely, and we denote them as \textit{Hawkes}.
The second set of features, denoted as \textit{Monsted}, replicates the work of Monsted et al~\cite{Monsted2018}, who use autoregressive models to account for the historical dependency between current and historical volumes of calls in a series. 
They also build call and text statistics, such as duration, inter-event time, or volume.
Unlike the original paper, we do not build features concerning online social network activity (Facebook), proximity (Bluetooth), and geospatial mobility (GPS) -- for which the data is not fully available in our studies.
The third set of features, denoted as \textit{Surveys}, builds upon the surveys filled by students.
For both \Netsense and \Nethealth, we extract information that describes the students, such as their grades, perception of their health, number of books they read, or the number of clubs or social organizations they belong to.
This results in 145 features for \Netsense and 2212 features for \Nethealth.
Note that \textit{Surveys} use the information provided by the students, whereas \textit{Hawkes} and \textit{Monsted} use only externally observed information (i.e., call series).
Note also that \textit{Surveys} does not contain the forty-four questions related to the Big5 questionnaire (and which serve as ground truth for the predictive exercise).
Finally, we also test combinations of feature sets to assess their complementarity.

\hypertarget{sec:inferring-psychometric-traits}{\subsection{Inferring psychometric traits}}

\label{subsec:traits}

\noindent\textbf{Predictive setup.}
We generate the three sets of features for each of the 178 users for \Netsense and 712 for \Nethealth for whom the \textit{Hawkes} features exist (see \hyperlink{sec:analyzing-call-patterns-using-hawkes}{Section \textit{Analyzing call patterns using Hawkes}}).
We trialed several off-the-shelf regressors -- including SVM, XGBoost, KNN, and Random Forests -- out of which Random Forests performed best and is shown in the rest of this section.
We tune the hyperparameters and compute prediction performance using 5-fold nested cross-validation and random search with 500 combinations.

\noindent\textbf{Performance of \textit{Hawkes} features.}
In line with recent literature~\cite{Kosinski2013, Stachl2020, Monsted2018,Novikov2021}, we report in \cref{fig:Big5-pearson} personality traits prediction performances using Pearson's correlation coefficient.
We show the RMSE performances in the online appendix~\cite{appendix}.
In their recent survey study, Novikov et al~\cite{Novikov2021} state that the upper limits for correlations between predicted and self-reported personality traits vary between 0.42 and 0.48.
Our \textit{Hawkes} features achieve 0.29 for \emph{openness} on \Netsense, and 0.37 for \emph{extraversion} on \Nethealth -- very close to the upper limit range.
The other traits are predicted around 0.17 on \Nethealth.
For \Netsense, \textit{Hawkes} performs poorly for \emph{conscientiousness}, \emph{agreebleness} and \emph{neuroticism}.

\noindent\textbf{\Netsense vs \Nethealth.}
We observe a higher variability of performance for \Netsense than for \Nethealth and a significantly higher variance.
We posit this is due to the length of the studies -- \Netsense it is only two and a half years, while for \Nethealth the period is four and a half years -- and the amount of data available for each.
Since personality traits do not change significantly over the years~\cite{Novikov2021}, longer communication periods may contribute to more accurate predictions.
We conclude that \Nethealth is a better dataset for predicting personality traits from communication data.

\noindent\textbf{Surveys features.}
Unsurprisingly, the best performing features overall are the \textit{Surveys} which embed self-reported information -- the gold standard in psychology.
On \Netsense, they achieve around 0.3, and for \Nethealth it obtains more than 0.7 on \emph{conscientiousness} and \emph{neuroticism}.
We make two observations.
First, \textit{Features} outperform the other feature sets given the richness of self-reported information (145 questions for \Netsense and 2212 for \Nethealth).
However, we note that surveying is a very costly process both financially and in time.
Second, \textit{Hawkes} and \textit{Surveys} appear to capture complementary aspects of personality, as putting them together further increases performances, particularly for \textit{openness} and \textit{extraversion} on both datasets.

\noindent\textbf{Hawkes vs Monsted.}
For \Nethealth, the \textit{Hawkes} features are outperforming the \textit{Monsted} on four out of the five traits and even come close to the prediction based on the very complex \textit{Surveys}.
The most significant advantage of \textit{Hawkes} over \textit{Monsted} is that it does not require complex feature engineering, which may not be easily portable to new datasets.
\hypertarget{sec:related-work}{\section{Related Work}}

We structure related works discussion into two sections.
First, we discuss work that used Hawkes processes to model contacts and interactions.
Second, we explore works that infer personality traits from online data sources.

\noindent\textbf{Modeling interactions in contact networks using point processes.}
Hawkes processes have been widely used to model social interactions in several applications because they can account for bursts of activity localized in time.
Zipkin et al~\cite{Zipkin2016} study electronic communications in a dataset of emails pertaining to the US military, where they observe that activity along the edges of the communication network is bursty.
They apply a Hawkes model for the email exchange along edges and focus on studying parameter estimation in the presence of missing data.
Moore and Davenport~\cite{moore2016analysis} learn the topology of a wireless network from limited passive observations of network activity.
They use a multi-variate Hawkes process;
they show it can detect changes to the existing topology and extract higher-level summaries of information flow in the network.
Choudhari et al~\cite{Choudhari2018} models simultaneously events and the structure of a social network using a Hidden Markov Hawkes Process that incorporates topical Markov Chains within Hawkes processes to jointly model topical interactions along with the user-user and user-topic patterns.
Hawkes processes have also been used to model face-to-face interactions in offices~\cite{masuda2013self} as well as retweet cascades.
For the latter task, Kobayashi and Lambiotte~\cite{Kobayashi2016} propose a Time-Dependent Hawkes process to account for the circadian nature of the users and the aging of information when modeling retweet cascades, whereas Mishra et al~\cite{Mishra2016FeaturePrediction} leverages a Hawkes process with a power-law relaxation kernel and uses it with the user and timing features to predict the popularity of retweet cascades. 
Luo et al~\cite{luo2019fused} use company employees' phone calls to study the alignment method for Hawkes processes based on fused Gromov-Wasserstein discrepancy.

To the best of our knowledge, this is the first work that applies Hawkes modeling to telecom data and uses its outputs to predict users' relations and personality traits.

\noindent\textbf{Predicting personal traits from social interactions.}
A fertile area of computational psychology deals with inferring psychometric traits from a range of sources made available by our new interconnected society.
For example, Settanni et al~\cite{Settanni2018} show that digital traces from social media can be studied to assess and predict theoretically distant psychosocial characteristics with remarkable accuracy. 
They also show that when additional user demographics are leveraged as additional digital traces, the accuracy of predictions improves.
Some of these works also use social media-inferred personality traits to influence opinions and behavior or infer private traits. 
For example, Matz et al~\cite{Matz2017} performed three field experiments that reached over 3.5 million individuals with psychologically tailored advertising and found that matching the content of persuasive appeals to individuals' psychological characteristics rendered the messaging significantly more effective.
Kosinski et al~\cite{Kosinski2013} used a public source of social media data (i.e., Facebook likes) to predict the Big5 personality traits of users, alongside a range of highly sensitive personal attributes, including sexual orientation, ethnicity, religious and political views, personality traits, intelligence, happiness, use of addictive substances, parental separation, age, and gender.
A follow-up study by Youyou et al~\cite{Youyou2015} even showed that personality traits prediction based on Facebook likes are more accurate than the user's friends' estimations based on surveys.

The prior work most relevant to this study relates to learning personality traits from behavioral information collected via smartphones.
Such works usually collect data from the onboard sensors and other phone logs~\cite{Harari2019}. 
The works closest to ours are by Stachl et al~\cite{Stachl2020}, and Monsted et al~\cite{Monsted2018}.
Stachl et al~\cite{Stachl2020} predict Big Five personality dimensions using six different classes of behavioral information collected from smartphones: 
1) communication and social behavior, 2) music consumption, 3)~app usage, 4) mobility, 5) overall phone activity, and 6) day- and night-time activity.
They find that the accuracy of these predictions is similar to that of social media platforms.
On the contrary, Monsted et al~\cite{Monsted2018} claim that smartphone usage is not as predictive of Big5 personality traits as previously reported, and higher predictabilities in the literature are likely due to overfitting on small datasets.

Our work differs from those mentioned above in two major aspects.
First, we use the Hawkes models fitted on the call series to predict user traits.
As far as we are aware, no other work has used fitted Hawkes point processes to distill the call interactions between users and predict Big5 traits.
Second, the above work requires access to the user's phone, as most features need to be recorded on the device. 
Our work shows that personality traits can be accurately predicted solely on call and text logs, which can be obtained outside the user's device.
We also show that our Hawkes descriptors' prediction accuracy is comparable to that obtained from user-filled surveys.

\section{Conclusion and Future Work}
This work investigates whether the Hawkes processes trained on telecommunication metadata predict human sociological traits and aspects. 
We employ two extensive datasets -- \Netsense and \Nethealth, which contain mobile phone communication metadata and detailed information on surveyed university students. 
We fit a Hawkes process for each pair (participating student, peer) and use its fitted parameters as descriptors. 
In a series of experiments, we show that the Hawkes processes can distinguish between types of relationships, predict their temporal dynamics and infer user Big5 psychometric traits. 
This work is the first to show that Hawkes processes can be used to abstract detailed communication events that carry additional human sociological information. Due to the novelty of our application of the Hawkes process in the described problem, we decided that the right approach would be to start research with a simple model, the vanilla Hawkes process. 
Our work and results could provide a starting point for future research that will focus on using more complex point process models.

\noindent\textbf{The ethics of using telecom traces to infer private traits.}
These capabilities raise critical questions concerning user privacy. 
Our work joins a body of evidence~\cite{Soto2011,Kosinski2013,Smith-Clarke2014} to show how the users' digital traits can be used to infer the user's sensitive information (such as the psychological profile).
It can be disconcerting that such rich sources of private information lie outside the control of users and in the data warehouses of third parties, and we argue that its usage should be regulated.
As Taylor~\cite{taylor2016no} argues, studying human mobility using telecom data is a double-edged sword.

\noindent\textbf{Limitations and future work.}
For a given pair of individuals (sender, receiver), this work merges outgoing and incoming calls into a single call series.
Future work could apply a bivariate Hawkes process to model the intertwining of both aspects of a discussion. In addition, as mentioned earlier, future work may focus on using more complex point process models. For example, call activity may be correlated with time, so it may be essential to use a model that takes into account the daily or weekly seasonality in the background rate~\cite{omi2017hawkes}. Another direction of research is to use deep learning, 
however, this will result in a loss of interpretability of the model.

\section{Data availability statement}
The \Netsense dataset is available upon request from Prof. Omar Lizardo, \Nethealth data is publicly available for research purposes: \url{https://sites.nd.edu/nethealth/}. The code and the data samples are available online at \url{https://github.com/pwr-ai/predicting-relationships-and-big5-using-hawkes}.

\bibliographystyle{plain}
\bibliography{biblio-new,bibliography}


\begin{IEEEbiography}[{\includegraphics[width=1in,height=1.25in,clip,keepaspectratio]{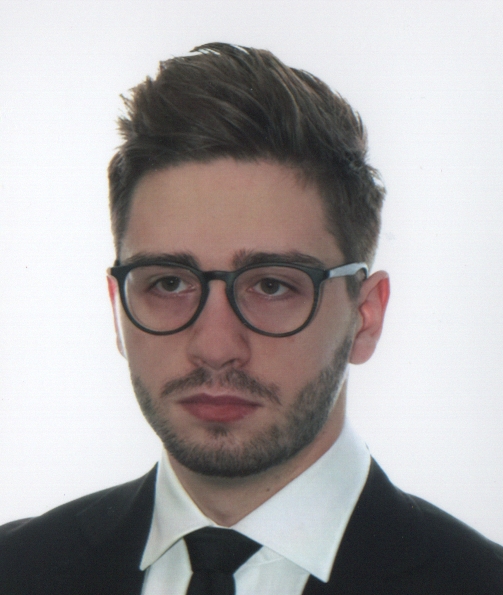}}]{Mateusz Nurek} is a PhD student at the \href{https://ai.pwr.edu.pl}{Department of Artificial Intelligence} at Wrocław University of Science and Technology in Wrocław, Poland and a member of the \href{https://networks.pwr.edu.pl}{Network Science Lab}. His research interests include network science and machine learning. He focuses on applying data science techniques in the field of human behavior and social network analysis.
\end{IEEEbiography}

\begin{IEEEbiography}[{\includegraphics[width=1in,height=1.25in,clip,keepaspectratio]{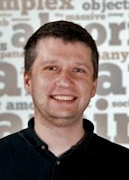}}]{RADOS{\L}AW MICHALSKI} is an Associate Professor at the \href{https://ai.pwr.edu.pl}{Department of Artificial Intelligence} at Wrocław University of Science and Technology (Poland). His research areas cover but are not limited to social influence, diffusion processes in complex networks, and machine learning. He has co-authored over 50 publications in these areas. He co-leads the \href{https://networks.pwr.edu.pl}{Network Science Lab} at Wrocław University of Science and Technology and leads \href{https://bergplace.org}{BERG - Blockchain Exploration Research Group}.
\end{IEEEbiography}

\begin{IEEEbiography}[{\includegraphics[width=1in,height=1.25in,clip,keepaspectratio]{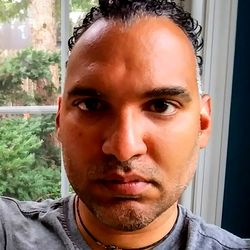}}]{Omar Lizardo}  is a Professor and LeRoy Neiman Term Chair at the University of California Los Angeles. He studies culture, cognition, networks, consumption, institutions, organization, and social theory. He currently serves on the editorial board of the social science journals \textit{Social Forces}, \textit{Sociological Theory}, \textit{Sociological Forum}, \textit{Journal for the Theory of Social Behaviour}, \textit{Theory and Society}, and \textit{Poetics}. He is also a member of the Board of Reviewing Editors for the journal \textit{Science} and an Associate Editor for \textit{Discover Data}.
\end{IEEEbiography}

\begin{IEEEbiography}[{\includegraphics[width=1in,height=1.25in,clip,keepaspectratio]{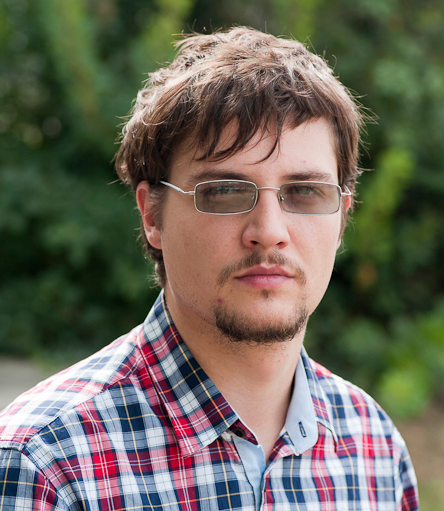}}]{Marian-Andrei Rizoiu} 
is an Assistant Professor with the University of Technology Sydney, where he leads the \href{https://www.behavioral-ds.science}{Behavioral Data Science} group. 
He is a computer scientist interested in addressing research questions where human online behavior and machine learning cross over. 
His research has made several key contributions to online popularity prediction, real-time tracking and countering disinformation campaigns, and understanding shortages and mismatches in labor markets.

Marian-Andrei's research is funded by Facebook Research and the Commonwealth of Australia, and published in the most selective venues, such as the PNAS, PLOS ONE, PLOS Computations Biology, WWW, NeurIPS, IJCAI, and CIKM. 
In addition, his work has received significant media attention, including \href{https://www.bloomberg.com/news/features/2020-02-12/the-best-way-to-change-your-job-focus-on-your-personality}{Bloomberg Business Week}, \href{https://www.natureindex.com/news-blog/scientists-are-curious-and-idealistic-but-not-very-agreeable-compared-to-other-professions}{Nature Index}, \href{https://www.bbc.com/worklife/article/20200123-how-your-twitter-feed-could-help-find-your-dream-job}{BBC}, and \href{https://www.weforum.org/agenda/2021/05/we-spent-six-years-scouring-billions-of-links-and-found-the-web-is-both-expanding-and-shrinking/}{World Economic Forum}. 
His work has a societal impact for social good, he serves as an expert in legislative initiatives and parliamentary inquiries. 
See more at \url{www.behavioral-ds.science}.
\end{IEEEbiography}

\EOD

\end{document}